\documentclass[twocolumn,amssymb]{revtex4}
\usepackage{graphicx}% Include figure files

\begin{document}

\title{Skyrmion Excitation in Two-Dimensional Spinor Bose-Einstein Condensate}
\author{ \it Hui Zhai, Wei Qiang Chen, Zhan Xu, and Lee Chang\\ Center for Advanced Study,
Tsinghua University, 100084 Beijing, China}
%\date{\today}
\begin{abstract}
We study the properties of coreless vortices(skyrmion) in spinor
Bose-Einstein condensate. We find that this excitation is always
energetically unstable, it always decays to an uniform spin
texture. We obtain the skyrmion energy as a function of its size
and position, a key quantity in understanding the decay process.
We also point out that the decay rate of a skyrmion with high
winding number will be slower. The interaction between skyrmions
and other excitation modes are also discussed.

\end{abstract}
\maketitle

\section{INTRODUCTION}
Topological objects have been attracting interest from various
fields of physics for several decades\cite{a}. Roughly speaking,
there are two kinds of topological excitations in two-dimensional
space. The configuration of the first kind, such as vortex or
monopole, has a natural singularity and depends on azimuthal
angles even at infinity. These excitations have infinite kinetic
energy unless they are coupled to other fields vanishing at
infinity. In order to obtain a topological structure with finite
energy, the configuration of the second kind must be uniform at
infinity, which means that the topological structure should be
defined in a compactified space. The skyrmion is an example of the
second kind topological structure, and the skyrmion excitation in
$n$-dimensional space exists when the $n$th homotopy group of the
internal space is nontrivial.

Since its introduction in the 1960s in nuclear physics\cite{b} and
its application in QCD\cite{c}, skyrmions have been found in
condensed matter systems such as quantum Hall effect\cite{d} and
high temperature superconductivity. The achievement of
Bose-Enistein condensate (BEc) in dilute Bose gases provide an
opportunity to investigate many-body theory in new system. So far
some topological excitations such as vortices and vortex rings in
scalar BEc have been observed in a number of labs\cite{e}.
Recently the realization of a spinor BEc\cite{f}, whose spin
degrees of freedom are unfrozen has generated interest in studying
richer topological excitations in such
systems\cite{g}\cite{h}\cite{i}.

The mean field description of spin-$1$ BEc was proposed in pioneer
works of Ho and Ohmi et al\cite{j}. The symmetry group of the
ground state order parameter of an antiferromagnetic spin-1 BEc is
found to be $U\left( 1\right) \times S^{2}/Z^2$,\cite{Zhou} where
$U(1)$ denotes the global phase angle and $S^2$ is a unit sphere
denoting all orientation of the spin quantization axis. The
additional $Z^2$ is because the refection of spin quantization
axis is equivalent to the change the global phase by $\pi$. The
general form of such a spinor field is

\begin{equation}
\zeta =\frac{1}{\sqrt{2}}\left(
\begin{array}{c}
-m_{x}+im_{y} \\
\sqrt{2}m_{z} \\
m_{x}+im_{y}
\end{array}
\right)
\end{equation}
where $\overrightarrow{m}$ is the Bloch vector. It should be
pointed out that the second homotopy group of $S^{2}$ is
homomorphic to integer group, this not only tells us of the
existence of monopole excitation in three-dimensional
antiferromagnetic BEc, which had been investigated by \textit{\
}H.T.C. Stoof\ \textit{\ et al}\cite{h}, but also implies the
existence of the skyrmion excitation in two-dimensional case.

It is naturally to ask if this excitation mode is energetically
stable or not. In this paper, we answer the question within the
Thomas-Fermi approximation and a variational method, and we find
that the skyrmion is always energetically unstable. In other
words, in presence of any energy dissipation mechanism, the spin
texture will always decay. However, this result does not imply
that the skyrmion can not be created, a skyrmion with half
topological charge has been successfully generated in a recent
experiment by adiabatic deformation of the magnetic trap\cite{ex}.
Although the half skyrmion created in this experiment is different
from that discussed in this paper, it is believed that a skyrmion
with integer winding number can also be created in the near
future.

Furthermore, it is useful to discover whether the skyrmion decays
by expanding to infinity size or shrinking to tiny size after it
is created, and to understand how its center of mass moves. To
answer these two questions, we need to obtain the energy of
skyrmion as a function of its size and position, this energy
function is the central result of our paper. We also find that the
presence of a vortex influences skyrmion's motion. Additionally
based on the results of Ref.\cite{g}, we will point out that the
decay rate of high winding number skyrmions will be much slower
than those with winding number $1$.

\section{The energetic stability of $Q=1$ Skyrmion}
When we only focus on the energy property of skyrmions, we can
first neglect the interaction in spin channel which disrupt the
$S^2$ order parameter, and the energy functional for such BEc can
be simplified as following:

\begin{widetext}
\begin{eqnarray}
K\left( \varphi ,\zeta \right) =\int\int d^2\vec{r}\left(
\frac{\hbar^{2}}{2m}\left| \nabla \varphi \right|
^{2}+\frac{\hbar^{2}}{2m}\left| \nabla \zeta \right| ^{2}\left|
\varphi \right| ^{2}-\left( \mu
-V_{trap}(\overrightarrow{r})\right) \left| \varphi \right|
^{2}+\frac{4\pi\hbar^2 a_{sc}N}{m}\left| \varphi \right|
^{4}\right) \label{1}
\end{eqnarray}
\end{widetext}
$\mu $ is the chemical potential and $V_{trap}$ is the confining
trap potential. In this section, we will use conformal mapping to
construct a general skyrmion excitation with winding number $Q=1$.
These variational wave functions are used to show that such
skyrmions are always energetically unstable.

Equation (\ref{1}) is a nonlinear sigma model coupled to a
$\varphi ^{4}$ model in an external potential. An $n$-skyrmion is
an instanton solution to the $1+1$ dimensional nonlinear sigma
model\cite{a}, this skyrmion minimizes the energy functional in
the sector of each homotopy class, and can be constructed from the
help of fractional linear mapping which maps the compactified
complex space into itself, and the most general form of this
mapping is

\begin{equation}
\Omega
=f(z)=\prod\limits_{i=1}^{N}\frac{a_{i}z+b_{i}}{c_{i}z+d_{i}}
\label{conformal}
\end{equation}
with the constraint $a_{i}d_{i}-b_{i}c_{i}=1$, where $N$ is the
winding number. Complex number $\omega$ are mapped to vectors on
the unit sphere via

\begin{equation}
m_{x}=\frac{\Omega +\bar{\Omega }}{1+\left| \Omega \right| ^{2}}%
,m_{y}=-i\frac{\Omega -\bar{\Omega }}{1+\left| \Omega \right| ^{2}}%
,m_{z}=\frac{\left| \Omega \right| ^{2}-1}{\left| \Omega \right|
^{2}+1}
\end{equation}
Owing to the conformal invariance of the nonlinear sigma model,
its classical solutions are infinitely degenerate and the total
energy is
independent of the parameters in the analytical function $f\left( z\right) $%
. The coupling between the spinor field $\zeta $ and the
superfluid field $\varphi $ breaks the conformal invariance, and
skyrmions are not classical solutions to our model . The spin
texture defined by equation(\ref{conformal}) can however be used
in a variational calculation. The texture of spinor field
contributes an effective potential, which changes the density
profile. Given an effective potential, the minimum of the energy
functional, is function of the parameters in the conformal mapping
$f\left( z\right) $. This function will tell us the information
about the energetic stability of skyrmion excitation.

We first consider the $Q=1$ case, \ $\Omega =f\left( z\right) =\frac{az+b}{%
cz+d}$. It is not difficult to show that

\begin{equation}
\left| \nabla \zeta \right| ^{2}=8\frac{\left| f^{\prime }\left(
z\right)
\right| ^{2}}{\left[ 1+\left| f\left( z\right) \right| ^{2}\right] ^{2}}=%
\frac{8}{\left[ \left| cz+d\right| ^{2}+\left| az+b\right|
^{2}\right] ^{2}}
\end{equation}
On the condition $ad-bc=1$, the effective potential is
non-singular implying the density remains finite at the skyrmion
core. This result is consistent with the coreless character of
skyrmion excitation. The
effective potential has two barrier localized at $z=-\frac{d}{c}$ and $z=-%
\frac{b}{a}$, with the height $\left| c\right| ^{4}$ and $\left|
a\right| ^{4}$ respectively, where the density of the condensate
should be smaller than the surrounding. Recall that in the
repulsive case the interaction constant $g$ is positive, both the
interaction energy and the kinetic energy favor homogenous density
profile, too much undulation will no doubt increase the energy, so
the choice $c=0$ helps to decrease the energy. This value of $c$
also produces the most symmetric texture. The function $f\left(
z\right) $ is reduced to $a^{2}(z-z_{1})$, where the parameters
$a$ and $z_{1}$ characterizes the size and location of the
skyrmion. \cite{l}.

For simplicity, we first force $z_{1}=0$ to be at the center of
the trap. Non-zero $z_{1}$ will be discussed later. The effective
potential from the spin texture now is $\frac{1}{\left[ \left|
\frac{1}{a}\right| ^{2}+\left| a\right| ^{2}r^{2}\right] ^{2}}$,
and the full potential is showed in Figure 1 for different $a$.
Notice that the barrier height of the effective potential is
proportional to $\left| a\right| ^{4}$, and that integrating the
potential energy over the whole two-dimension space results in a
constant. In the limit $a\rightarrow 0$, the effective potential
spreads uniformly throughout the whole space. In the contrary
limit of infinite $a$, the effective potential becomes a $\delta $
function.

We now investigate the energetic stability of skyrmion in the
framework of TF approximation where the term $|\nabla\varphi|^2$
is neglected. In this approximation, the density profile is
\begin{widetext}
\begin{equation}
n\left( \overrightarrow{r}\right) =\left( \mu
-a_{HO}^{2}r^{2}-\frac{8}{\left[ \left( \frac{1}{a}\right)
^{2}+a^{2}r^{2}\right] ^{2}}\right) \frac{1}{2g}\Theta \left( \mu
-a_{HO}^{2}r^{2}-\frac{8}{\left[ \left( \frac{1}{a}\right)
^{2}+a^{2}r^{2}\right] ^{2}}\right) \label{2}
\end{equation}
\end{widetext}
in which $a_{HO}=\left( \frac{m\omega }{\hbar}\right) ^{2}$ and
$g=8\pi a_{sc}N$, $\omega $ is the frequency of the harmonic trap,
and $\Theta(x)$ is a step function. The normalization condition
and the constraint that the condensate density is non-negative
give the following two equations:

\begin{equation}
\frac{1}{2g}\int\limits_{0}^{x}\int\limits_{0}^{2\pi }rdrd\theta
\left[ \mu
-a_{HO}^{2}r^{2}-\frac{8}{\left[ \left( \frac{1}{a}\right) ^{2}+a^{2}r^{2}%
\right] ^{2}}\right] =1  \label{3}
\end{equation}

\begin{equation}
\mu -a_{HO}^{2}x^{2}-\frac{8}{\left[ \left( \frac{1}{a}\right)
^{2}+a^{2}x^{2}\right] ^{2}}=0  \label{4}
\end{equation}%
where $x$ is the value of $r$ at which the Thomas-Fermi density
vanishes. Solving the two equations we obtain the relationship
between the chemical potential $\mu $, Thomas-Fermi radius $x$ and
the skyrmion's size $a$. One can substitute these relationships
back to the energy functional and then obtain the minimal energy
$E$ as a function of size $a$ which is plotted in Figure 2.

Figure 2 shows that the minimal energy $E$ has two minima
occurring at zero and infinity, and there exists a critical size
$a_{c}$ which corresponds to the maximal value of $E$. When $a$ is
quite large, the TF approximation
fails, we use variational density profile to obtain a more accurate result\cite%
{m}, which is shown in Figure 3.

Physically, we can understand the two figures in the following
way. When the size parameter $a$ is sufficiently small, the spin
configuration within the TF radius becomes uniform, the effective
potential becomes flat inside the TF radius and just becomes a
uniform shift of the chemical potential. So the energy approaches
the ground state energy as $a\rightarrow0$. In the opposite limit,
the barrier is quite high and thus the density almost vanishes at
the center of the skyrmion, but at the same time the width of the
barrier becomes narrow. The energy cost of the low density region
is proportional to the volume and decreases as $a$ becoming
larger. In this case the skyrmion can be observable by imaging the
density profile, and observing the low density core. These figures
tell us that the $Q=1$ skyrmion is energetically unstable, in
presence of any weak dissipation, it will either expand to become
unobservable, or shrink to an infinitesimal size. As a
semi-classical object the dissipative dynamics of a skyrmion
depends on whether or not its initial size is larger than the
critical size $a_{c}$.

For the same reason, the position $z_{1}$ of the skyrmion tends to
move toward the edge of the condensate where the density is lower
and the energy cost $\int V_{eff}|\varphi|^2$ is smaller.

\begin{figure}[htbp]
\begin{center}
\includegraphics[width=2.1in]
{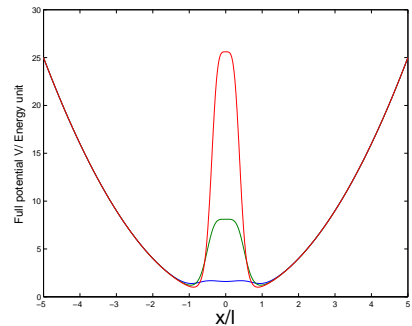} \caption{The trap potential added with the
effective potential induced by $Q=1$ skyrmion, in the unit of
$\hbar\omega$, is plotted as a function of $x/l$  for
$a/\sqrt{a_{HO}}=1,1.5,2$ respectively. Here $l=1/\sqrt{a_{HO}}$}
\end{center}
\end{figure}

\begin{figure}[htbp]
\begin{center}
\includegraphics[width=2.1in]%
{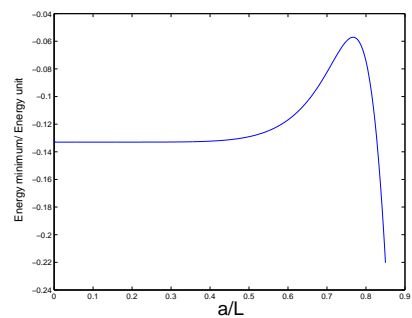}%
\caption {The minimal energy $E$ in the unit of $\hbar\omega$ vs.
the size of skyrmion $a$ in the unit $L=\sqrt{a_{HO}}$. The
parameter $g/a_{HO}$ is set to $2$. }
\end{center}
\end{figure}

\begin{figure}[htbp]
\begin{center}
\includegraphics[width=2.1in]%
{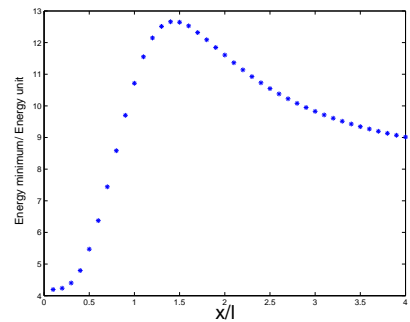}%
\caption{The minimal energy $E$ in the unit of $\hbar\omega$ vs.
size $a$ in the unit of $L$, the result is obtained using
variational method.}
\end{center}
\end{figure}
\section{Dissipation Dynamics}
Although we believe that the energetically unstable skyrmion will
decay in presence of any weak dissipation, the exact description
of the dynamics depends on the details of the dissipation
mechanism. A possible source of dissipation is interaction with
the non-condensed component. Since skyrmion is a global
topological object and the system is in a quite low temperature,
this may not be important. Another source of dissipation is the
spontaneous emission of other excitation modes, such as phonons
mode or vortices, similar to the spontaneous emission of an
excited atom. This mechanism is observed in numerical studies of
superfluid vortex recconnection \cite{n}. Let us discuss the
process in more detail.\ When $a$ increases, the peak of effective
potential grows and atoms are pushed away from the center of the
skyrmion. The kinetic energy of these atoms will lead to
oscillatory motion of the condensate density. If the trap harmonic
potential is anisotropic, the extracted atoms can also rotate to
form a vortex. These modes take energy away from the skyrmion,
leading to a change of the skyrmion size.

\section{Interaction Between A Vortex and A Skyrmion}

We rewrite $\varphi $ as $\Psi e^{i\theta }$. In equation
(\ref{1}) the phase field $\theta $ is not directly coupled to the
spinor field $\zeta$, but they both interacts with the density
field. Since both vortices and skyrmion have low density cores, it
is favorable that the skyrmion core fits entirely within the
vortex, thus one expects that the presence of vortex change the
dominant trend of the evolvement of the skyrmion from expanding to
shrinking, and one should find the skyrmion attracted to the
vortex center instead of moving toward the cloud's edge.

In principle, we can integrate out the $\Psi $ field in the action
and obtain the effective Hamiltonian to describe the interaction
between a vortex and a skyrmion. Unfortunately the $\Psi^4$ term
makes this procedure difficult. However in the TF approximation,
where the derivative terms of $\Psi$ are neglected, the action is
a quadratic form of the density field $n$. We can integrate it out
and obtain an effective Hamiltonian describing the interaction
between $\theta$ and $\zeta$:
\begin{equation}
H_{int}=-\frac{1}{2g}\left(\frac{\hbar^2}{2m}\right)^2\int\int d^2
\vec{r}\left( \left| \nabla \theta \right| ^{2}\left| \nabla \zeta
\right| ^{2}\right)
\end{equation}
The interaction energy will be smallest when the maximum of $
\left| \nabla \theta \right| $ and $\left| \nabla \zeta \right|$
coincide. This result is consistent with the above picture.

\section{$Q>1$ Skyrmion}
In this section we discuss skyrmions with $Q>1$, demonstrating
their differences from the $Q=1$ skyrmion. Following the logic of
the second section, we choose the most symmetric case, $\ f\left(
z\right) =a^{2}z^{n}$,
finding the effective potential
\begin{equation}
V_{n}=8\frac{ n^{2}r^{2n-2}}{[(\frac{1}{a})^{2}+a^{2}r^{2n}]^{2}}
\end{equation}
The barrier of $V_{n}$
($n>2$) lies on a circle
\begin{eqnarray}
r=\left( \frac{2n-2}{2n+2}\left( \frac{1}{a}%
\right) ^{4}\right) ^{\frac{1}{2n}}
\end{eqnarray}
with the height
\begin{eqnarray}
V_{max}=2\left( n^{2}-1\right) \left(
\frac{2n+1}{2n-1}a^{4}\right) ^{\frac{1}{n}}
\end{eqnarray}
this structure is markedly different from $V_{1}$ whose peak is
always localized at its center. In the center, the effective
potentials is always zero for any $n\geq 2$. Figure 4 shows
$V_{2}$ for different value of $a$. As $a$ is decreasing, the
location of the off-center peak of the potential approaches
infinity as $\left( \frac{1}{a}\right) ^{\frac{4}{2n}}$ and its
height decreases as $a^{\frac{4}{n}}$. Because the size of
condensate is always finite, this implies that it is energetically
favorable for a skyrmion to increase to a larger size so that the
peak of the effective potential is outside the TF radius and the
effective potential becomes small and uniform inside the
condensate. What is different from the $Q=1$ case is the atoms
must tunnel through the barrier of the effective potential as the
size increases. Thus, the rate of the process will be
characterized by the WKB tunnelling rate. The situation has been
studied carefully in Ref\cite{g}, and it showed that the
tunnelling rate of such process may be long enough to form a
dynamic metastable skyrmion.

\begin{figure}[htbp]
\begin{center}
\includegraphics[width=2.1in]%
{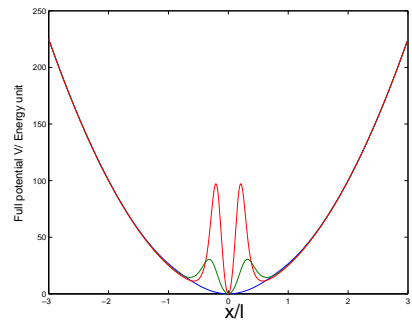}%
\caption {The trap potential added with the effective potential
induced by $Q=2$ skyrmion, in the unit of $\hbar\omega$, is
plotted as a function of $x/l$ for $a/\sqrt{a_{HO}}=1,3$ and $5$
respectively}
\end{center}
\end{figure}

\section{Conclusion}
In summary, we constructed and studied the skyrmion excitation
mode of the spinor field and showed that any finite size skyrmion
is always unstable. Through a variational study of the skyrmion
excitation energy for different size and position, we find that it
is energetically favorable for skyrmion to expand to infinite size
or shrinking to infinitesimal size, resolution in a uniform spin
texture in most part of the condensate. We discussed the interplay
between skyrmion modes and other excitation modes: (1) the
emission phonons can dissipate the skyrmion energy and change its
size, and (2) the vortex has attractive interactions with a
skyrmion. We also showed that the behavior of $Q>1$ skyrmions are
quite different from $Q=1$ skyrmion since the effective potentials
induced by their configuration have some markedly difference.
However, all these discussions are concentrated on properties of
the skyrmion mode its itself and a qualitative prediction on the
skyrmion dynamics is made from the energetic consideration. To
study the detail of skyrmion dynamic process, it is important to
consider the interaction in spin channel, which cause spin
fluctuations exceeding the $S^2$ internal space.\cite{h}

In the end, noticing that $\zeta $ represents local spin-gauge
degrees of freedom, we remark that the model we explored in this
paper shares some features of Yang-Mills theory, that is, the
instanton solutions having the same winding number are saddle
points of the Langrangian and are energetically degenerate
although they have different shape and size, because of the local
scaling invariance of the pure gauge theory. However, when the
coupling to matter field breaks the conformal symmetry, the
instanton solutions are no longer classical solutions and have
different actions. In addition, the dynamic term of gauge field is
absent in our model. The conclusion of energetic instability of
any finite size skyrmion is determined by the above two features.

\textit{Acknowledgements}: The authors thank Professor C.N. Yang
for encouraging. We acknowledge Professor Tin-Lun Ho for helpful
discussion, and Doctor Erich Mueller, L$\ddot{u}$ Rong for his a
lot of wonderful comments and suggestions. This work is supported
by National Natural Science Foundation of China ( Grant No.
10247002 and 90103004 )


\begin{thebibliography}{99}
\bibitem{m} The trial wave function we used here takes the form $\left(
a+br^{2}\right) e^{-\omega ^{2}r^{2}}$.
\end{thebibliography}
\end{document}